\documentstyle[epsfig,amssymb,aps,multicol]{revtex}

\begin{document}

\draft
 
\title{Electrostatically swollen lamellar stacks and adiabatic pair
potential of highly charged plate-like colloids in an electrolyte}

\author{Emmanuel Trizac
\footnote{Electronic Address: Emmanuel.Trizac@th.u-psud.fr}}

\address{
Laboratoire de Physique Th\'eorique (Unit\'e Mixte de Recherche UMR 8627 du CNRS), \\
B\^atiment 210, Universit\'e de Paris-Sud, 91405 Orsay Cedex, France
}
 
\date{\today}

\maketitle
\begin{abstract}
We show that describing the screened electrostatic 
interactions in a periodic stack of rigid
parallel identical charged plate-like colloids within a local density functional theory
approach, generically leads to a swelling behaviour. Within the same framework, 
we find that the effective potential between a pair of such plates
immersed in an electrolyte is repulsive at all distances. 
This general result is in contradiction with a theory put
forward by Sogami, Shinohara and Smalley, that we criticize and 
thoroughly reconsider to show 
how the two approaches can be reconciled. 
\end{abstract}



\section{Introduction}
The counter-intuitive phenomenon of effective attractions between like-charges
immersed in an electrolyte has recently attracted considerable attention,
and questions one of the fundamental tenets of colloid science
\cite{Kepler,Gronbech,Rene,Warren,Brenner}. 
The understanding of the underlying mechanism is crucial for a correct
description of a vast variety of industrial and natural processes, in particular the
stability and phase behaviour of colloidal dispersions. In this article, we investigate
the stability of electrostatically swollen stacks of regularly spaced planar colloids
or membranes of infinite lateral extension in an electrolyte. 
Such a one-dimensional geometry 
describes the early stages of clay swelling \cite{vanOlphen} but is also relevant
for lamellar phases of charged bilayers \cite{Dubois}.
At the simplest level of mean-field description, with 
neglect of excluded volume and Coulomb correlation effects,
the inhomogeneous density profiles of
microscopic co- and counter-ions forming the electric double layers around the 
planar macroions can be obtained analytically \cite{Gouy}. The 
resulting Poisson-Boltzmann osmotic (or disjoining) pressure in the stack, is  
invariably found positive \cite{Andelman}, which is the signature of the tendency
to swell. 
This behaviour is reminiscent of the repulsive nature of pair interactions
(as we shall see below, the two phenomena are however distinct),
established within the same mean-field framework \cite{Neu,Sader}
and recently extended beyond mean-field \cite{Papier}, with the possible 
inclusion of approximate correlation contributions.
Note that more refined incorporation of discrete solvent effects by 
adding a bilinear non local term to the free energy of
standard Poisson-Boltzmann theory has shown the possibility
of a net attraction (negative pressure) at high surface charges 
\cite{Andelmanbis}.

The purpose of the present investigation is twofold. First, 
we apply the general local density functional 
formalism of Ref. \cite{Papier} to provide a prediction for the stability
of stacked or lamellar phases. We show in section 
\ref{sec:swelling} that the interactions
of electric double layers generically lead to swelling whereas describing 
an equilibrium spacing between the membranes either requires the inclusion
of non electrostatic forces (usually van-der-Waals like)
or the consideration of more refined theories.
Second, the pair potential problem is considered in section III
with a resulting effective repulsion at all distances, within the same framework
that encompasses in particular the non-linear Poisson-Boltzmann theory.
This statement is consequently in contradiction with the attractive behaviour
reported by Sogami, Shinohara and Smalley \cite{Sogami1,Sogami2}. We show that their
controversial finding is attributable to a confusion in the thermodynamic 
potential describing the electrostatic situation under study.

\section{Swelling of a lamellar stack}
\label{sec:swelling}
Before setting the framework of the analysis, it is worthwhile to point out 
that the results obtained in \cite{Neu,Sader,Papier} for a pair of colloids 
do not apply to the regular stacked situation under scrutiny 
here (the precise definition of the geometry is given 
at the beginning of section IIA). 
First of all, the mechanical route followed in Refs. 
\cite{Neu,Sader,Papier} (deriving the 
interactions by integration of the stress tensor over the colloids' surfaces)
would yield
a vanishing effective force in a regular stack, with cancelling contributions 
acting on both sides of the platelets, and is thus uninformative about the stability
of the array. Moreover, the situation considered in \cite{Neu,Sader,Papier}
is that of two colloids immersed in an electrolyte, confined in a cylinder
with an axis parallel to the colloids' line of centers, and of infinite extension 
along this axis [see Figure \ref{figure}-a)]. The assumption of infinite length is
crucial for the validity of the effective repulsion 
(see Appendix \ref{app:a}) and the
corresponding confined two-body problem does not include the multi-body
stack geometry, for which the confining cylinder (Wigner-Seitz cell)
would be a slab of finite length. This point is in contradiction with one 
erroneous conclusion reached in Ref. \cite{Raimbault}.

\subsection{Density functional theory formalism}
We consider a regular succession of rigid infinite parallel plates located at 
$z=2 n h$ ($n\in\mathbb{N}$). Each plate carries a uniform surface charge
$\sigma$,  and releases its counter-ions in the electrolyte solution,
considered to be a mixture of $N$ microions, where species $\alpha$ has 
charge number $e_\alpha$ and local density $n_\alpha(z)$. We write the free
energy\cite{Hansen} of the total charge distribution in the Wigner-Seitz cell around a 
given platelet (e.g. that situated at $z=0$ for which the cell is a slab extending
between $-h$ and $h$) as
\begin{equation}
{\cal F}(\{n_\alpha\}) = \int_{-h}^h f[\{n_\alpha(z)\}]\,d z \,+\, 
\frac{1}{2} \, \int_{-h}^h \rho_c(z)\,\psi(z) \,  d z,
\label{eq:lda}
\end{equation}
where $\rho_c(z)$ is the local total charge distribution, including the
microions and the plate (with global electroneutrality 
$\smash{\int_{-h}^h \rho_c = 0}$). 
${\cal F}$ is expressed per unit area of the macroscopic surfaces. 
Within the primitive model of electrolytes \cite{Hansenrevue}, 
whereby the solvent is regarded as a mere dielectric continuum of permittivity
$\varepsilon$,
the total electrostatic potential $\psi(z)$ is
the solution of Poisson's equation
\begin{equation}
\nabla^2 \psi\,=\,-\frac{4\pi}{\varepsilon} \,\rho_c(z) \,=\,
-\frac{4\pi}{\varepsilon} \, \left[\sigma \delta(z)\,+\,
\sum_{\alpha} e_\alpha\,n_\alpha(z)
\right]
\label{eq:poisson}
\end{equation}
and can be cast in the form $\psi(z) = \int_{-h}^h \rho_c(z')\,G(z,z')\,dz'$ where
$G$ is the appropriate Green's function. The boundary condition of vanishing 
electric field at $z=\pm h$ is fulfilled by $\partial_z \psi$. This mean-field
like reduction of the problem materialized by 
the introduction of the Wigner-Seitz slab,
may be corrected by correlation terms included in the free energy density $f$.
Moreover, even if the van der Waals energy term 
$\int \rho_c \psi = \int \rho_c G \rho_c$ is of mean-field form, 
correlation terms can be accounted for provided they translate into a local 
correction to the free energy, 
in the spirit of the approaches depicted in Refs \cite{Stevens,Lowen,Barbosa}.
Consequently, the term $\int f$ in Equation (\ref{eq:lda})
does not reduce in general to the entropic
microions' contribution (as in Poisson-Boltzmann theory \cite{Hansenrevue}) but 
may also
include both non mean-field energetic and entropic effects. Of course, the present
formalism encompasses the non-linear Poisson-Boltzmann and Modified 
Poisson-Boltzmann theories \cite{Eigen,Lue}.

\subsection{Osmotic pressure}
\label{ssec:osmotic}
For both canonical (fixed ionic content of the electrolyte solution) and
semi grand-canonical (when the solution is in osmotic equilibrium with 
a salt reservoir) descriptions, the optimal density profiles $n_\alpha^*(z)$
are obtained by minimizing the functional
\begin{equation}
\Omega(\{n_\alpha\}) \,=\,{\cal F}(\{n_\alpha\})\,-\,\sum_\alpha \mu_\alpha N_\alpha;
\qquad \left(N_\alpha \,=\,\int_{-h}^h n_\alpha\, dz\right),
\end{equation}
where $\mu_\alpha$ is either the Lagrange multiplier associated with the constraint of
fixed average concentration, or the chemical potential of species $\alpha$ in the
reservoir. The resulting stationary equations read for all species $\alpha$
\begin{equation}
\frac{\partial f}{\partial n_\alpha} \,+\, e_\alpha \psi(z) \, = 
\mu_\alpha.
\label{eq:station}
\end{equation}
These implicit relations between the electrostatic potential and the densities
\{$n_\alpha$\} allow to solve Poisson's equation (\ref{eq:poisson}) 
and compute the optimal 
profile $n_\alpha^*(z)$ (hereafter noted $n_\alpha(z)$ without ambiguity), from which
we deduce the Helmholtz free energy $F(h,\sigma,T,\{N_{\alpha}\}) = 
{\cal F}(\{n_\alpha^*\})$, with $F=U-TS$ ($U$ is the total internal energy,
$T$ the absolute temperature and $S$ the entropy of the total micro-ions
charge distribution
in the cell). 

From the knowledge of $F$, the definition of the thermodynamic potential
${\cal R}$ (which variations correspond to the reversible work performed
by an operator and thus define the osmotic pressure) requires the specification
of the thermodynamic situation under consideration. In the canonical 
case with constant charge plates, ${\cal R} = F$ \cite{Landau}. If on the other
hand, the platelets are held at constant potential (e.g. by an external generator),
${\cal R}$ is the Legendre transform of $F$ with respect to the 
surface charge \cite{Landau,Callen}, namely 
\begin{equation}{\cal R}=F(\sigma,h,T)-\sigma
\frac{\partial F}{\partial \sigma}\biggl|_{h,T}
=F(\sigma,h,T)-\sigma \psi_{_{\cal P}}(\sigma,h,T),
\end{equation}
where $\psi_{_{\cal P}}=\psi(z=0)$ denotes the surface potential.
In the opposite case of a system in equilibrium with a salt reservoir, 
\begin{eqnarray}
&&{\cal R}\, =\, {\cal R}_{\sigma}\, =\,\Omega(\{n_\alpha\}) \,
=\,F\,-\,\sum_\alpha \mu_\alpha \int_{-h}^h n_\alpha(z)\,dz 
\qquad \hbox{at constant charge $\sigma$}\\
&&{\cal R}\, =\, {\cal R}_{\psi}\, =\,F\,-\,\sum_\alpha \mu_\alpha 
\int_{-h}^h n_\alpha(z)\,dz\,- 
\,\sigma\frac{\partial F}{\partial \sigma}
\qquad \hbox{~~~~at constant potential $\psi_{_{\cal P}}$}.
\label{eq:constantpot}
\end{eqnarray}
In any case, the osmotic pressure is defined as
\begin{equation}
\Pi \,=\, -\frac{1}{2}\,\frac{\partial {\cal R}}{\partial h},
\end{equation}
where ${\cal R}$ depends on the electrostatic situation considered,
as explained above. It is however important to stress that the osmotic
pressure should not depend on the electrostatic situation under study,
as becomes clear below. Theories that do not result in the above fundamental
invariance of $\Pi$ can consequently be disposed of. 

The free energy variation induced by changing the inter-membrane distance is
computed in Appendix \ref{app:b} with the result
\begin{equation}
\delta F\,=\,\left[f-\sum_{\alpha} n_{\alpha}\frac{\partial f}{\partial n_\alpha} 
\right]_{z=h} 2 \delta h\,+\, \sum_{\alpha}\mu_\alpha\delta N_{\alpha}\,+\,
\psi_{_{\cal P}} \delta \sigma,
\label{eq:Fvar}
\end{equation} 
so that the osmotic pressure reads
\begin{equation}
\Pi \,=\, \left[-f+\sum_{\alpha} n_{\alpha}\frac{\partial f}{\partial n_\alpha} 
\right]_{z=h},
\end{equation}
independently of the situation of constant potential or constant charge considered.
Not surprisingly, the pressure is related to the Legendre transform of the
free energy density, as for ordinary homogeneous 
gases \cite{Callen} and equals the local stress
$\pi(z)$ at the mid-plane between the membranes:
\begin{equation}
\Pi\,=\pi(z=h)\quad \hbox{where} \quad \pi(z) \,=\,-f[\{n_\alpha\}(z)]+\sum_{\alpha} n_{\alpha}(z)
\frac{\partial f}{\partial n_\alpha}(z).
\label{eq:piofz}
\end{equation}

\subsection{Stability analysis}
\label{ssec:stability}
When the stack is in osmotic equilibrium with a salt reservoir, the 
comparison of $\Pi$ obtained in section \ref{ssec:osmotic} with the pressure
exerted by the reservoir quantifies the tendency towards swelling 
($\Pi >\Pi_{\hbox{\scriptsize res}}$) or collapse 
($\Pi <\Pi_{\hbox{\scriptsize res}}$). For consistency, the (neutral) reservoir
needs to be described within the same framework as the electrolyte around
the platelet. The remainder of this section is devoted to the proof that
$\Pi$ is extremal in the reservoir, and that this extremum is a minimum
under fairly general conditions. We first analyze the  $\psi$ 
dependence of the pressure $\pi$, defined by $\pi(\psi)= \pi(z)$ where $\psi=\psi(z)$
is the solution of Poisson's equation (\ref{eq:poisson}). 
From Eq. (\ref{eq:piofz}) we have
\begin{equation}
\frac{\partial \pi}{\partial \psi} \,=\, \sum_{\alpha} n_{\alpha}
\frac{\partial}{\partial \psi}\left(\frac{\partial f}{\partial n_\alpha}\right),
\end{equation}
that can be recast making use of the stationary condition (\ref{eq:station})
\begin{eqnarray}
\frac{\partial \pi}{\partial \psi} &=& -\sum_{\alpha=1}^N n_{\alpha} e_{\alpha} \\
&=& -\rho_c(z) \qquad \hbox{outside the plate $(z\ne 0)$}.
\label{eq:dpi}
\end{eqnarray}
$\pi$ thus goes through an extremum in the reservoir 
($\Pi_{\hbox{\scriptsize res}}$ by definition)
where the charge density
vanishes (unlike at the mid-plane $z=h$ where $\rho_c\ne 0$).
Relation (\ref{eq:dpi}) together with Poisson's equation implies that the
local stress introduced in (\ref{eq:piofz}) fulfills the mechanical
equilibrium condition for the fluid of microions:
\begin{equation}
\pi(z)\,-\,\frac{\varepsilon}{8 \pi} \left(\frac{\partial \psi}{\partial z}\right)^2
\,=\, \Pi,
\label{eq:equilmech}
\end{equation}
which is the simple form taken in a one-dimensional problem 
by the condition of vanishing divergence for the generalized Maxwell stress
tensor in the dielectric medium \cite{Landau}. 

The second derivative of $\pi$ can be obtained by introducing the Hessian
matrix $N\times N$
\begin{equation}
H_{\alpha\beta} \,=\, \frac{\partial^2 f}{\partial n_\alpha \partial n_\beta}
\end{equation}
that obeys the relation
\begin{equation}
\sum_{\beta=1}^N H_{\alpha \beta}
\frac{\partial n_\beta}{\partial \psi}\,=\,- e_{\alpha}
\end{equation}
as can be seen from Eq. (\ref{eq:station}). We thus have
\begin{equation}
\frac{\partial^2 \pi}{\partial \psi^2} \,=\,
\sum_{\alpha \beta} H_{\alpha \beta}
\frac{\partial n_\alpha}{\partial \psi}
\frac{\partial n_\beta}{\partial \psi},
\end{equation}
which is a positive quantity when the matrix of second derivatives $H_{\alpha\beta}$
is positive definite. Under this assumption, $\pi$ is a convex-up 
function of the potential and the extremum attained in the reservoir 
is therefore a minimum, so that $\Pi >\Pi_{\hbox{\scriptsize res}}$.

In spite of the differences outlined at the beginning of section \ref{sec:swelling}
between the pair potential and the multi-body stacked problems,
the method employed here is close to that of Ref. \cite{Papier},
and we shall also distinguish two different situations to conclude with 
the stability analysis
\begin{itemize}
\itemsep=0pt
\item {\em Case a)}: the free energy density does not depend on the elementary charge
$e$ (as in mere mean-field treatments \cite{Gouy,Hansenrevue,Eigen,Lue}).
In the limit $e\to 0$ (at fixed valency $e_\alpha/e$), we obtain a locally
neutral mixture where the total free energy functional reduces to the 
contribution $\int f$. The thermodynamic stability criterion of this neutral
mixture implies the positive definiteness of the matrix $H_{\alpha\beta}$
that is independent on $e$. 
\item {\em Case b)}: correlation or fluctuation effects are taken into account
with a resulting $e$-dependent free energy density. It is no longer possible 
to find an uncharged mixture of micro-species described by the same  
density $f$. The thermodynamic stability condition of the full functional
(\ref{eq:lda}) involves the Green's function $G(z,z')$ and does not imply
the positive definiteness of the Hessian $H$. However, the 
convexity of $f$ with respect to density variations is generally fulfilled
by the approaches proposed in the literature, either in the full density
range \cite{Barbosa} or for the small plasma
coupling parameters relevant to  colloidal dispersions \cite{Lowen}. 
\end{itemize}
From the above discussion and the positive definiteness of the Hessian matrix $H$, we
conclude that the stack generically exhibits a tendency towards swelling.

\section{Adiabatic pair potential}

In this section, we consider within the density functional formalism of Eq. (\ref{eq:lda}), 
the problem of the effective interactions
between a pair of parallel charged plates immersed in an electrolyte solution of
infinite volume (no confinement). The two rigid plates with distance $2h$ 
are supposed to be of negligible
thickness and uniform surface charge (as in section \ref{sec:swelling});
they divide the electrolyte solution into two disconnected regions
(inner region with $|z| < h$ and outer region with $|z|>h$), which does
not correspond to the situation analyzed in Refs. \cite{Neu,Sader,Papier}.
However, the generalization to the present case is straightforward 
and yields an effective repulsion as long as both regions are in contact 
with salt reservoirs imposing the same chemical potential for 
micro-species. Indeed, from the computation of the free energy variation 
detailed in section \ref{ssec:osmotic}, the effective force can be written
\begin{equation}
F_z \,=\, -\frac{1}{2}\,\frac{\delta {\cal R}}{\delta h}\,=\,
\Pi_{\hbox{\scriptsize in}} - \Pi_{\hbox{\scriptsize out}}.
\label{eq:force}
\end{equation}
From the vanishing of the electric field at $z=0$ and $|z| \to \infty$, 
we have $\Pi_{\hbox{\scriptsize in}} = \pi(z=0)$ and 
$\Pi_{\hbox{\scriptsize out}} = \pi(|z| \to\infty)$. 
For $|z|\to \infty$, the charge density vanishes so that 
$\Pi_{\hbox{\scriptsize out}}$ equals the osmotic pressure in the salt
reservoir under consideration. The argument of section
\ref{ssec:stability} indicates that 
$\Pi_{\hbox{\scriptsize in}} >\Pi_{\hbox{\scriptsize out}}$ and that 
the interactions are repulsive under the assumption of positive definiteness
for the stability matrix \smash{$\partial^2_{\alpha \beta} f$}. 
This last condition is 
obeyed by Poisson-Boltzmann (PB) theory independently of the valency of 
the microions [see Eq. (\ref{eq:freepb}) below]. 
Our result is consequently in contradiction with 
the ``long-range weak attractive part of the
free energy'' reported in \cite{Sogami1,Sogami2} for the same system
treated at the level of Poisson-Boltzmann. The work of Sogami
{\it et al.}\/ has already been criticized 
\cite{Levine,Ettelaie,Overbeek,Ruckenstein},
but in our opinion, the subsequent controversy \cite{Smalley}
dwells on ambiguities on the thermodynamic potential that should 
be considered, which to our knowledge, have not been explicitly pointed out
so far. It thus seems worthwhile to restrict to PB theory
and devote the remainder of this article to briefly revisit the model of 
\cite{Sogami1,Sogami2}, introduced to describe the swelling behaviour of
n-butyl-ammonium vermiculite gels. 

Within PB mean-field theory, the microions are considered as
an ideal gas and density fluctuations discarded, so that the free
energy density does not include any correlation term and reduces to the
entropy of an ideal mixture:
\begin{equation}
f(\{n_\alpha\})\,=\, \beta^{-1}\,\sum_{\alpha=1}^N n_\alpha\left[
\ln\left(n_{\alpha} \Lambda_\alpha^3\right) -1\right],
\label{eq:freepb}
\end{equation}
where the (irrelevant) lengths $\{\Lambda_\alpha\}$ involve the masses of microions
and $\beta = 1/(kT)$ is the inverse temperature. The stationary condition 
(\ref{eq:station}) translates into 
\begin{equation}
n_\alpha(z) = n_\alpha^0 \exp(-\beta e_\alpha \psi)
\label{eq:density}
\end{equation}
and the local osmotic stress in
(\ref{eq:piofz}) is given by the ideal equation of state
$\pi(z) = kT\sum_\alpha n_\alpha(z)$. The electrostatic potential 
$\psi$ is chosen to vanish for $|z| \to \infty$, so that 
$n_\alpha^0$ is the density of species $\alpha$ far from the plates 
with a corresponding chemical potential
\begin{equation}
\mu_\alpha = kT \ln\left(n_{\alpha}^0 \Lambda_\alpha^3\right).
\label{eq:mualpha}
\end{equation}
As a result of the global electroneutrality condition, it 
can be checked that the final free energy [expression (\ref{eq:freepbmod}) below]
is independent on an arbitrary shift
of the potential $\psi$ [in which case the densities appearing in Eq.
(\ref{eq:mualpha}) are simply the prefactors of the exponentials in (\ref{eq:density})].
It is convenient to use the relation between the local ionic 
densities and the electrostatic
potential to recast the Helmholtz free energy $F=U-TS$ in the form:
\begin{eqnarray}
F \,=\, {\cal F}(\{n_\alpha\}) &\!=\!& 
\sigma \psi_{_{\cal P}} -\frac{\varepsilon}{8 \pi}\,
\int_{-\infty}^{\infty} \left(\frac{\partial \psi}{\partial z}\right)^2 \, dz
\,+\,\sum_{\alpha} N_\alpha \left(\mu_\alpha -kT\right).
\nonumber\\
&=& \sigma \psi_{_{\cal P}} + \int_{-\infty}^{\infty} \Pi \, dz\, + \,
\sum_{\alpha} N_\alpha \left(\mu_\alpha-2 kT\right),
\label{eq:freepbmod}
\end{eqnarray}
where Eq. (\ref{eq:equilmech}) and the ideal equation of state
for $\pi(z)$ have been used in going from the first to second
line. Note that it is understood that the osmotic term $\Pi$ in
(\ref{eq:freepbmod}) takes the value $\Pi_{\hbox{\scriptsize in}}$ 
(resp. $\Pi_{\hbox{\scriptsize out}}$) for $|z|<h$ (resp. $|z|>h$).
Strictly speaking, expression (\ref{eq:freepbmod}) diverges
(if salt is added to the electrolyte, some of the quantities
$N_\alpha$ are extensive with system size). This feature can be circumvented
by computing the excess free energy with respect to a well chosen reference 
system (for instance the system with same bulk densities in the absence of the plates). 
Once the solution of Poisson's equation is known (see \cite{Andelman} for
a review of the standard solutions, including the present geometry), 
$F$ may be computed from equation (\ref{eq:freepbmod}). 
Following this route, we readily recover the Helmholtz
free energy obtained by Sogami {\it et al.} by means of a charging process.
In Equation (\ref{eq:freepbmod}), 
$\sigma=\sigma_{\hbox{\scriptsize in}} + \sigma_{\hbox{\scriptsize out}}$ 
stands for the total
surface charge on a platelet, including both the different
inner ($\sigma_{\hbox{\scriptsize in}}$ from $z=h^-$) and outer 
($\sigma_{\hbox{\scriptsize out}}$ from $z=h^+$) 
contributions, denoted respectively 
$Z_i$ and $Z_o$ in \cite{Sogami1,Sogami2}.

In the model of \cite{Sogami1,Sogami2}, the electrostatic potential is 
imposed to be continuous throughout the system and the surface 
potential $\psi_{_{\cal P}}$ is independent on the distance $2h$ between the 
plates. Moreover, the numbers of microions between the plates
also depend on $h$ with a fixed chemical potential, given by
(\ref{eq:mualpha}) where the $n_{\alpha}^0$ can be considered as the ionic densities
of a reservoir in chemical equilibrium with both the inner and outer
parts of the electrolyte solution around the plates.
Consequently, the thermodynamic potential ${\cal R}_\psi$
defined in Eq. (\ref{eq:constantpot}) should be used
in computing the force whereas $F$ has been considered in \cite{Sogami1,Sogami2}.
The definition of the model imposes that when $h$ changes at fixed 
$\psi_{_{\cal P}}$, $Z_o$ is constant while $Z_i$ varies, so that the two
situations of constant charge and constant potential are equivalent
for the outer part of the system ($|z|>h$), but not for the inner part.
The reversible work performed by an operator changing the distance $2h$ between
the charged plates is therefore given by the variations of
\begin{equation}
{\cal R}_\psi \,=\, F_{\hbox{\scriptsize in}} + F_{\hbox{\scriptsize out}}\,
-\, \sum_\alpha \mu_\alpha \int_{-\infty}^{\infty} n_{\alpha}(z)\, dz\,-\,
\sigma_{\hbox{\scriptsize in}} \psi_{_{\cal P}},
\label{eq:potpot}
\end{equation}
whereas $F=F_{\hbox{\scriptsize in}} + F_{\hbox{\scriptsize out}}$ has been
considered in \cite{Sogami1,Sogami2}. Omission of the chemical potential terms
$\mu_\alpha n_\alpha$ in the right hand-side of Eq. (\ref{eq:potpot})
leads the force computed in \cite{Sogami1,Sogami2} to depend on the masses
(through the lengths $\Lambda_\alpha$), which is impossible in an equilibrium
statistical mechanics theory, as already pointed out in \cite{Kjellander}.
With the aid of the relations given in \cite{Sogami1} [e.g. Eqs (63), (64)
and (65) valid in the case of counter-ion dominance between the plates],
an explicit computation of the force $ F_z = -\partial {\cal R}_\psi/\partial (2h)$
yields the standard expression
\begin{eqnarray}
F_z &=& \pi(h^-) -\frac{\varepsilon}{8 \pi} 
\left(\frac{\partial \psi}{\partial z}\right)^2_{z=h^-} -
\left[\pi(h^+) -\frac{\varepsilon}{8 \pi} 
\left(\frac{\partial \psi}{\partial z}\right)^2_{z=h^+}\right]
\label{eq:potea}\\
&=& \frac{\varepsilon}{8 \pi} \left[
\left(\frac{\partial \psi}{\partial z}\right)^2_{z=h^+}-
\left(\frac{\partial \psi}{\partial z}\right)^2_{z=h^-}
\right],
\label{eq:poteb}
\end{eqnarray}
where the continuity of the co- and counter-ion charge density 
across the membrane (resulting from the imposed continuity of the potential)
has been used in going from (\ref{eq:potea}) to (\ref{eq:poteb}).
The compatibility of Eq. (\ref{eq:potea}) with (\ref{eq:force}) is transparent.
In terms of the variables used in \cite{Sogami1,Sogami2}, where
$\Phi(0)$ denotes the reduced potential at mid-distance between the plates,
we get
\begin{equation}
F_z\,=\, \frac{2\pi}{\varepsilon}\,e^2 \left(Z_o^2-Z_i^2\right)\,=\,
4 n_0 \,kT \,\hbox{sinh}^2\left[\frac{\Phi(0)}{2}\right] \,\geq 0,
\end{equation}
hence an effective repulsion at all distances, that is in the present
case entirely due to the electrostatic pressure (the osmotic contribution
cancelling on both sides of the plates).

It can be checked that in the dual situation where both the inner
$\sigma_{\hbox{\scriptsize in}}$ are outer $\sigma_{\hbox{\scriptsize out}}$
surface charges are held constant, the effective force is still given by
(\ref{eq:potea}). This translates into the Legendre identity
\begin{equation}
\frac{\partial {\cal R}_\psi}{\partial h}
\biggl|_{\psi_{_{\cal P}}\,,\sigma_{\hbox{\scriptsize out}}}\, =\,
\frac{\partial}{\partial h}\left[{\cal R}_\psi - \psi_{_{\cal P}}
\frac{\partial {\cal R}_\psi}{\partial \psi_{_{\cal P}}}
\biggl|_{\psi_{_{\cal P}}\,,\sigma_{\hbox{\scriptsize out}}} 
\right]_{\sigma_{\hbox{\scriptsize in}}\,,\,\sigma_{\hbox{\scriptsize out}}}
\,=\, \frac{\partial \left(
{\cal R}_\psi +\sigma_{\hbox{\scriptsize in}} \psi_{_{\cal P}}\right)}
{\partial h}\biggl|_{\sigma_{\hbox{\scriptsize in}}\,,\,\sigma_{\hbox{\scriptsize out}}}
\label{eq:identity}
\end{equation}
that can be considered as a test for the consistency of the thermodynamic
potential used. 
In the limit of counter-ion dominance between the plates investigated
in \cite{Sogami1}, the identity
(\ref{eq:identity}) can be checked explicitly with the thermodynamic
potential $\cal R_\psi$ used here.

\section{Conclusion}

Describing the interactions between the electric double-layers around
charged planar colloids with a local density functional theory for ionic
screening and the primitive model of electrolytes, we have shown that a regular
stack of such plates generally displays a swelling behaviour when 
electrostatic forces alone are taken into account. Within the same
framework, the effective pair potential is found to be repulsive at all
distances, in contradistinction with the results derived in 
\cite{Sogami1,Sogami2}. The repulsive interactions are evidenced
without resort to an explicit solution of Poisson's equation,
and are related to the convexity of the underlying free energy
functional, as already noted in \cite{Papier}. 

Of course, the present result does not preclude the possibility of
effective attractive pair potentials. In particular, 
a drawback of the theories encompassed by Eq. (\ref{eq:lda}) is that the
direct correlation function $c^{(2)}({\bf r}_1,{\bf r}_2)$ defined as the second
functional derivative of ${\cal F}$ is necessarily mean-field like:
\begin{equation}
c_{\alpha,\beta}^{(2)}({\bf r}_1,{\bf r}_2)\,=\,
-\frac{\partial^2 f}{\partial n_\alpha \partial n_\beta}\delta({\bf r}_1-{\bf r}_2)
\,-\,e_\alpha e_\beta G({\bf r}_1,{\bf r}_2),
\end{equation}
with a $\delta$-correlated short range part. This shortcoming may be circumvented
by the inclusion of non-local terms in the theory, e.g. in the spirit of
the weighted density approximation \cite{Diehl,Groot}. However, we expect
the local formalism considered here to be instructive for inter-plate
separations much larger than the ionic size.

{\appendix
\section{}
\label{app:a}
References \cite{Neu,Sader,Papier}  proved that the effective interactions between
a pair of like-charged colloids immersed with counter-ions and salt in a 
confining cylinder of infinite extension [see Figure \ref{figure}-a)] 
were repulsive. We show here that the result does not hold for a finite 
length cylinder.

We consider two colloids confined in the cylinder of length $2h$ represented in
Figure \ref{figure}-b), in the specific case of Neumann boundary conditions
on the surface $\Sigma$ (vanishing normal electric field). 
Due to the mirror symmetry of the charge distribution, the electric field has
no $z$-component in the plane $z=0$, and the problem is equivalent to that of
a unique colloid (say colloid 1) in a sub-cell cylinder of length $h$, 
with again Neumann boundary conditions [the right half part of the cylinder,
indicated by a dotted rectangle in Fig. \ref{figure}-b)]. 
We assume the effective force acting on colloid 1 repulsive ($F_1^z\geq 0$).
Then, we translate along the $z$ axis the left half cell ($-h\leq z\leq0$) 
by a distance $2h$, keeping the right half fixed. The electrostatic potential around colloid 1 is unaffected,
so that the effective force is unchanged, corresponding now to an effective
attraction. Any repulsive configuration with Neumann boundary conditions
on the confining cylinder can then be mapped onto an attractive one 
(from the construction of the mapping, it appears that this feature disappears
in the limit $h \to \infty$).
The hypothesis of infinite length is thus a key ingredient
of the proofs in Refs. \cite{Neu,Sader,Papier}. 
It is moreover worthwhile to note that numerical solutions of the non-linear 
Poisson-Boltzmann equation for finite-size disc-like clay platelets
confined in a finite length Wigner-Seitz cylinder, show repulsive
effective pair forces when the distance between the clay particles is smaller 
than the half length $h$ of the cylinder \cite{Raul}.
\begin{figure}
\null\vspace{-0.5cm}$$\input{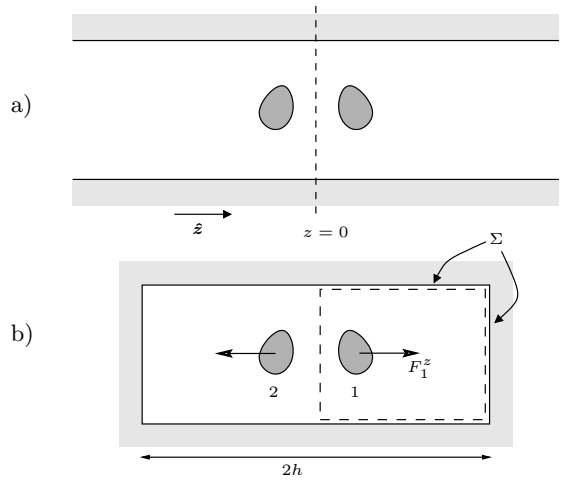}$$
\caption{  
Illustrative side view of the cell models considered. Mirror symmetry
with respect to the mid-plane $z=0$ between the two colloids is enforced
(this requirement is an important ingredient in the proofs 
\protect\cite{Neu,Sader,Papier}).}
\label{figure}
\end{figure}

\section{}
\label{app:b}
It will be shown that upon modifying the distance between the 
plates ($h \to h + \delta h$),
the Helmholtz free energy changes according to Eq. (\ref{eq:Fvar}). The 
present derivation bears some similarities with other ones in the related
context of Poisson-Boltzmann cell theory \cite{Marcus}. From the definition of the
free energy, Equation (\ref{eq:lda}) and the symmetry $z\leftrightarrow -z$, 
we have
\begin{equation}
\delta F \,=\, \int_{-h}^h \delta f\, dz \,+\,2\int_h^{h+\delta h} f\, dz\,+\,
\frac{1}{2}\,\int_{-h}^h \delta(\rho_c\psi)\, dz \,+\,\int_{h}^{h+\delta h}
\rho_c \psi\, dz.
\end{equation}
Consider first the energetic contribution. Making use of Poisson's equation
(\ref{eq:poisson}), two integrations by parts [with vanishing terms
$(\partial_z \psi)^2_{z=\pm h}$] yield
\begin{eqnarray}
\frac{1}{2}\,\int_{-h}^h \delta(\rho_c\psi)\, dz \!&=&\!
\int_{-h}^h \psi \delta \rho_c \, dz \,+\,
\int_{h}^{h+\delta h} \rho_c \psi\, dz  \label{eq:appb1}\\
&=& \int_{-h}^h \psi \delta \rho_c\,dz \,+\, \sum_{\alpha=1}^N \int_{h}^{h+\delta h}
n_\alpha\left(\mu_\alpha - \frac{\partial f}{\partial n_\alpha}\right) dz.
\label{eq:appb2}
\end{eqnarray}
The stationary condition (\ref{eq:station})
was used in going from (\ref{eq:appb1}) to (\ref{eq:appb2}). Similarly, with
$N_\alpha = \int_{-h}^h n_\alpha\,dz$ and since $\rho_c$ reduces to the 
microions charge distribution $\sum_\alpha e_\alpha n_\alpha$ outside the
platelets (in particular between $h$ and $h+\delta h$)
\begin{equation}
\int_{-h}^h \delta f\, dz \,=\, \sum_{\alpha=1}^N \int_{-h}^h \left(
\mu_\alpha - e_\alpha\psi\right)\delta n_\alpha\, dz \,=\,
\sum_{\alpha=1}^N \mu_\alpha \left(\delta N_\alpha - 2 \int_{h}^{h+\delta h}
n_\alpha \right)\,-\,\int_{-h}^h \psi \delta \rho_c\, dz + \psi(z=0)\delta \sigma.
\end{equation}
Gathering results, we obtain equation (\ref{eq:Fvar}).

}



\end{document}